\documentclass[sn-mathphys-num]{sn-jnl}

\usepackage{graphicx}%
\usepackage{multirow}%
\usepackage{amsmath,amssymb,amsfonts}%
\usepackage{amsthm}%
\usepackage{mathrsfs}%
\usepackage[title]{appendix}%
\usepackage{xcolor}%
\usepackage{textcomp}%
\usepackage{manyfoot}%
\usepackage{booktabs}%
\usepackage{algorithm}%
\usepackage{algorithmicx}%
\usepackage{algpseudocode}%
\usepackage{listings}%
\usepackage{adjustbox}
\usepackage{txfonts}

\theoremstyle{thmstyleone}%
\theoremstyle{thmstyletwo}%

\theoremstyle{thmstylethree}%

\begin{document}

\title[STDP]{Spike-timing-dependent plasticity and random inputs shape interspike interval regularity of model STN neurons}


\author*[1]{\fnm{Thoa} \sur{Thieu}}\email{thoa.thieu@utrgv.edu}

\author[2]{\fnm{Roderick} \sur{Melnik}}\email{rmelnik@wlu.ca}
\equalcont{These authors contributed equally to this work.}


\affil*[1]{\orgdiv{School of Mathematical and Statistical Sciences}, \orgname{The University of Texas Rio Grande Valley}, \orgaddress{\street{1201 W University Dr}, \city{Edinburg}, \postcode{78539}, \state{Texas}, \country{USA}}}

\affil[2]{\orgdiv{MS2Discovery Interdisciplinary Research Institute}, \orgname{Wilfrid Laurier University}, \orgaddress{\street{75 University Ave W}, \city{Waterloo}, \postcode{N2L 3C5}, \state{Ontario}, \country{Canada}}}



\abstract{Neuronal oscillations are closely related to the symptoms of Parkinson's disease (PD). In this study, we explore how random fluctuations (or "stochastic inputs") affect these oscillations in brain states, which reflect the collective activity of interconnected neurons. These random inputs are modeled in the context of the subthalamic nucleus (STN), a brain region implicated in PD, and their interaction with synaptic dynamics and spike-timing-dependent plasticity (STDP) in both healthy and PD-affected neurons. Specifically, we investigate the effects of random synaptic inputs and their correlations on the membrane potential of STN neurons. Our results show that these random inputs significantly influence the firing patterns of STN neurons, both in healthy cells and in those affected by PD under deep brain stimulation (DBS) treatment. We also find that STDP increases the regularity of the interspike intervals (ISI) in spike trains of output neurons. However, the introduction of random refractory periods and fluctuating input currents can induce greater irregularity in the spike trains. Furthermore, when random inputs and STDP are combined, the correlation between the activity of different neurons increases. These findings suggest that the stochastic dynamics of STN neurons, in conjunction with STDP, could offer insights into the mechanisms underlying PD symptoms and their potential management.}




\maketitle

\section{Introduction}\label{sec1}

Parkinson's disease (PD) is a progressive neurodegenerative disorder primarily characterized by the loss of dopaminergic neurons in the substantia nigra pars compacta, which leads to debilitating motor symptoms such as bradykinesia, rigidity, and tremors. Although various treatments, including pharmacological interventions and deep brain stimulation (DBS) of the subthalamic nucleus (STN), have shown efficacy in alleviating these symptoms, the mechanisms underlying these therapies remain poorly understood. Recent studies have highlighted the role of abnormal neuronal oscillations in the STN in contributing to motor dysfunction in PD, and suggested that synaptic plasticity-particularly spike-timing-dependent plasticity (STDP)-may play a crucial role in the modulation of these oscillations and the therapeutic effects of DBS \cite{Hirschmann2022,David2020, Rosa2011}.	

STDP is a form of synaptic plasticity in which the strength of a synapse is adjusted based on the relative timing of pre- and postsynaptic spikes. This form of learning has been shown to be critical for maintaining neural network stability and promoting coordinated activity \cite{Madadi2022, Wang2022}. In healthy brains, STDP is essential for processes like learning, memory, and sensory processing, but in the context of PD, its dysfunction may contribute to the altered neuronal activity observed in the STN \cite{Hirschmann2022, Lee2022}. Moreover, the role of STDP in PD pathophysiology remains underexplored, especially with respect to how random factors-such as fluctuations in synaptic inputs or noise from sensory and thermal sources-affect neural firing patterns and network dynamics \cite{Groen2019, Faisal2008}.	
This gap in understanding motivates our study, which aims to investigate how STDP interacts with stochastic factors in the dynamics of STN neurons. We develop a stochastic model that incorporates both random input currents and synaptic correlation, focusing on their combined effects on the cell membrane potential and neuronal spiking patterns in PD. While several studies have explored the effects of STDP and random inputs in neuronal networks, including modeling studies of the STN \cite{Ebert2014coordinated}, as well as studies investigating the mechanistic features of networks with STDP and noise \cite{Popovych2013self,Lucken2016noise}, the specific interaction between these factors in the context of STN neurons remains underexplored. Additionally, related studies such as those by \cite{Nobukawa2016enhancement,Baroni2010spike} further elucidate the role of noise and plasticity in network dynamics. Burkitt et al. \cite{Burkitt2004spike} also provide foundational insights into the general mechanisms of STDP in noisy environments, which are relevant for understanding the broader context of our study. Finally, additional work by Lindner \cite{Lindner2006superposition} offers valuable perspectives on noise-driven network dynamics that complement our investigation. Specifically, we explore how STDP influences the regularity of interspike intervals (ISIs) and how random factors such as fluctuations in synaptic inputs or noise from sensory and thermal sources, neural firing patterns and network dynamics \cite{Zheng2021, Powanwe2021}. This approach is distinct from prior studies, which typically examine STDP in more controlled, deterministic settings \cite{Thieu2022aims, Wang2022}.	

Our model builds on well-established frameworks, including the Hodgkin-Huxley model for single-neuron dynamics \cite{Hodgkin1990quantitative} and integrate-and-fire (LIF) models for simplifying neuron spiking behavior \cite{Izhikevich2003simple}, but uniquely integrates random input fluctuations and STDP to more accurately reflect the stochastic nature of biological neurons. Several studies have explored the interplay between noise and STDP in Hodgkin-Huxley (HH)-type neuron models \cite{Goldwyn2011and} and in leaky integrate-and-fire (LIF) models (e.g., \cite{Song2000competitive}). These studies provide important insights into how noise and plasticity together shape neuronal dynamics and learning processes. While previous work has focused on the effects of STDP in healthy networks, our study is one of the first to examine how STDP and random noise interact in the context of PD, a disease characterized by dysregulated neural circuits and irregular neuronal firing \cite{Sherf2020, Bujan2015}. Several studies have explored neural networks with STDP and noise in the context of PD. Chauhan et al. (2024) \cite{Chauhan2024synaptic} investigated how synaptic weight and structural plasticity co-evolve, revealing how structural reorganization can enhance network synchrony and inform desynchronization strategies for PD. Madadi Asl et al. (2018) \cite{Madadi2018propagation} showed that incorporating propagation delays and noise in STDP networks allows for bidirectional synapse preservation, offering insights into synaptic dynamics relevant to PD. These studies emphasize the importance of both excitatory and inhibitory inputs, which are crucial for accurately modeling the STN in PD. We build upon the deterministic STN model developed in \cite{So2012}, which was used to study the contributions of local cell and passing fiber activation to thalamic fidelity during DBS and lesioning. While their model considered deterministic dynamics, we introduce a stochastic term to capture the inherent random fluctuations in neuronal activity. These fluctuations, which arise from sensory noise, brainstem discharges, and thermal energy, are crucial for understanding the irregular firing patterns and synaptic plasticity observed in conditions like PD. By extending the model with this stochastic component, we investigate how random inputs influence the neuronal behavior and the efficacy of therapies like DBS.

The main contribution of this work is twofold. First, we propose a novel computational framework that combines STDP with random inputs to model the effects of noise and synaptic plasticity in PD. Second, we explore how these dynamics affect the regularity of neuronal firing, with potential implications for both PD pathophysiology and therapeutic interventions such as DBS. Our results suggest that while STDP can enhance the regularity of interspike intervals (ISIs), the presence of random input currents and refractory periods can introduce significant irregularity in spike trains, which may contribute to the motor symptoms observed in PD \cite{Wang2022, Kim2020}. These findings provide new insights into the neural mechanisms of PD and the potential role of STDP in managing the disease, with implications for future neuromorphic systems designed to mimic these processes \cite{Thieu2022-1, Ranganathan2018}.

\section{Model description}

STDP is a fundamental
mechanism in the brain that modifies the synaptic strengths
between neurons based on the coincidence of pre- and
postsynaptic spikes. In conventional asymmetric forms of
STDP, the temporal order of spikes is critical,
so that when the presynaptic spike precedes the postsynaptic
spike (i.e., pre-post pairing), the STDP rule leads to long-term
potentiation (LTP) of the synapse between pre- and postsynaptic
neurons, whereas long-term depression (LTD) is induced in the
reverse scenario (i.e., post-pre pairing).
Although STDP is a local mechanism and merely depends on the
pre- and postsynaptic spike timings, it can
determine global connectivity patterns emerging in recurrent
neuronal networks.

We will focus on building a model of a synapse in which its synaptic strength changes as a function of the relative timing (i.e., time difference) between the spikes of the presynaptic and postsynaptic neurons, respectively. This change in the synaptic weight is known as STDP. The aims of this paper are to build a model of synapse that show STDP and study how correlations in input spike trains influence the distribution of synaptic weights. We will model the presynaptic input as Poisson-type spike trains. The postsynaptic neuron will be modeled as an HH neuron.

We assume that a single postsynaptic neuron is driven by $N$ presynaptic neurons. That is, there are $N$ synapses, and we will study how their weights depend on the statistics or the input spike trains and their timing with respect to the spikes of the postsynaptic neuron.

The phenomenology of STDP is typically described by a biphasic exponentially decaying function and order-dependent, with pre-post spiking inducing LTP and LTD \cite{Markram1997regulation,Bi1998synaptic}. The instantaneous change in synaptic weight, $\Delta W$, is given by:
\begin{align}
	\Delta W &= A_{+}e^{(t_{\text{pre}} - t_{\text{post}})/\tau_{+}} \text{ if }  t_{\text{post}} > t_{\text{pre}},\\
	\Delta W &= -A_{-}e^{-(t_{\text{pre}} - t_{\text{post}})/\tau_{-}} \text{ if }  t_{\text{post}} < t_{\text{pre}},
\end{align}
where $\Delta W$ represents the change in synaptic weight, and $A_{+}$ and $A_{-}$ determine the maximum extent of synaptic modification. These maximum changes occur when the timing difference between presynaptic and postsynaptic spikes is close to zero. The parameters $\tau_{+}$ and $\tau_{-}$ define the time windows of presynaptic-to-postsynaptic interspike intervals that lead to synaptic strengthening or weakening \cite{Vilimelis2020spiking}. The latency between presynaptic and postsynaptic spikes ($\Delta t$) is defined as:
\begin{align} \Delta t = t_{\text{pre}} - t_{\text{post}}, \end{align} where $t_{\text{pre}}$ and $t_{\text{post}}$ represent the times of the presynaptic and postsynaptic spikes, respectively.  When the postsynaptic neuron fires after the presynaptic neuron (i.e., $\Delta t<0$), this leads to a positive change in synaptic weight ($\Delta W > 0$).

This model captures the key phenomenon that repeated presynaptic spikes occurring a few milliseconds before postsynaptic action potentials induce long-term potentiation (LTP) of the synapse, whereas presynaptic spikes occurring after postsynaptic spikes result in long-term depression (LTD) at the synapse.

%
%
%
%

\subsection{Keeping track of pre- and postsynaptic spikes}

To implement STDP while considering different synapses, we must track the times of both presynaptic and postsynaptic spikes throughout the simulation. To achieve this, we introduce two variables to monitor the spike times: \( M(t) \) for the postsynaptic neuron and \( P(t) \) for the presynaptic neuron. These variables are used to track the spikes over a longer timescale than the synaptic conductances.
For each postsynaptic neuron, we define the following differential equation to track the postsynaptic spike history (see, e.g., \cite{Song2000competitive}):
\[
\tau_{-}\frac{dM}{dt} = -M,
\]
where \( \tau_{-} \) is the timescale over which the postsynaptic spike influence decays. Whenever the postsynaptic neuron spikes, we update \( M(t) \) as follows:
\begin{align}\label{post}
	M(t) = M(t) - A_{-}.
\end{align}
This allows \( M(t) \) to track the number of postsynaptic spikes over the timescale \( \tau_{-} \), and ensures that \( M(t) \) is always negative, which is associated with the induction of LTD.
Similarly, for each presynaptic neuron, we define the following differential equation:
\[
\tau_{+}\frac{dP}{dt} = -P,
\]
where \( \tau_{+} \) is the timescale over which the presynaptic spike influence decays. When the presynaptic neuron spikes, we update \( P(t) \) as follows:
\begin{align}\label{pre}
	P(t) = P(t) + A_{+}.
\end{align}
Here, \( P(t) \) is always positive, corresponding to LTP. 
The variables \( M(t) \) and \( P(t) \) play similar roles to synaptic conductances \( g_i(t) \), but instead of controlling the synaptic strength, they track the temporal relationship between presynaptic and postsynaptic spikes over much longer timescales. Importantly, \( M(t) \) is negative and associated with LTD, while \( P(t) \) is positive and associated with LTP, since they are updated by \( A_{-} \) and \( A_{+} \), respectively.

\subsection{Implementation of STDP}
The peak synaptic conductance of each synapse \( i \), denoted by \( \bar{g}_i \), will be adjusted based on the timing of presynaptic and postsynaptic spikes, using the variables \( M(t) \) and \( P(t) \). The conductance \( \bar{g}_i \) can vary between \([0, \bar{g}_\text{max}]\) and is modified according to the relative timing of the spikes \cite{Song2000competitive}.
\begin{itemize}
	\item[1. ] When the 
	i-th presynaptic neuron fires a spike, the peak conductance is updated as follows:
	\begin{align}\label{g_M} \bar{g}_i = \bar{g}_i + M(t) \bar{g}_{\text{max}}. \end{align}
	Here, $M(t)$ tracks the time since the last postsynaptic spike and is always negative. Therefore, if the postsynaptic neuron spikes shortly before the presynaptic neuron, the peak conductance will decrease, as indicated by the negative value of $M(t)$.
	\item[2. ] When the postsynaptic neuron fires a spike, the peak conductance of each synapse is updated as follows:
	\begin{align}\label{g_P}
		\bar{g}_i = \bar{g}_i + P(t)\bar{g}_{\text{max}}. 
	\end{align}
	Here, 
	$P(t)$ tracks the time since the last spike of the 
	i-th presynaptic neuron and is always positive.
	
	Thus, if the presynaptic neuron spikes before the postsynaptic neuron, the peak conductance increases, as indicated by the positive value of 
	$P(t)$.
	%
\end{itemize}
\subsection{The leaky integrate-and-fire neuron connected with synapses that show STDP}

We connect $N$ presynaptic neurons to a single postsynaptic neuron. We do not need to simulate the dynamics of each presynaptic neuron as we are only concerned about their spike times. So, we will generate $N$ Poisson-type spikes.

We need to simulate the dynamics of the postsynaptic neuron as we do not know its spike times. We model the postsynaptic neuron as Hodgkin-Huxley (HH) system modelling an STN cell membrane potential.

Furthermore, motivated by \cite{So2012,Yang2014-n,Chen2018,Zhang2020}, we consider a modified HH system modelling a STN cell membrane potential. In particular, we first choose a STN healthy cell, then switch to a PD cell, and study the effects of random inputs on the STN cell membrane potential under synaptic conductance dynamics. 


In biological systems of brain networks, instead of physically joined neurons, a spike in the presynaptic cell causes a release of a chemical, or a neurotransmitter. Neurotransmitters are released from synaptic vesicles into a small space between the neurons called the synaptic cleft \cite{Gerstner2014}. In what follows, we will investigate the  chemical synaptic transmission and study how excitation and inhibition affect the patterns in the neurons' spiking output in our HH model.

In this section, we consider a HH model of synaptic conductance dynamics. In particular, neurons receive a myriad of excitatory and inhibitory synaptic inputs at dendrites. To better understand the mechanisms of synaptic conductance dynamics, we use the description of Poissonian trains to investigate the dynamics of the random excitatory (E) and inhibitory (I) inputs to a neuron \cite{Dayan2005,Li2019}. 

We consider the transmitter-activated ion channels as an explicitly time-dependent conductivity $(g_{\text{syn}}(t))$. The conductance transients can be defined by the following equation (see, e.g., \cite{Dayan2005,Gerstner2014}): 

\begin{align}\label{conductivity}
	\frac{d g_{\text{syn}}(t)}{dt} = -\bar{g}_i\sum_{k}\delta(t-t_k) - \frac{g_{\text{syn}}(t)}{\tau_{\text{syn}}},
\end{align}
where $\bar{g}_{\text{syn}}$ (synaptic weight) denotes the maximum conductance elicited by each incoming spike, while $\tau_{\text{syn}}$ is the synaptic time constant, and $\delta(\cdot)$ is the Dirac delta function. Note that the summation runs over all spikes received by the neuron at time $t_k$.  We have the following formula for converting conductance changes to the current by using Ohm's law:

\begin{align}
	I_{\text{syn}}(t) = g_{\text{syn}}(t)(V(t) - E_{\text{syn}}), 
\end{align}
where $V$ is the membrane potential, while $E_{\text{syn}}$ represents the direction of current flow and the excitatory or inhibitory nature of the synapse.


The total synaptic input current $I_{\text{syn}}$ is the combination of both excitatory and inhibitory inputs. Assume that the total excitatory and inhibitory conductances received at time $t$ are $g_E(t)$ and $g_I(t)$, and their corresponding reversal potentials are $E_E$ and $E_I$, respectively. Then, the total synaptic current can be defined by the following equation (see, e.g., \cite{Li2019}): 

\begin{align}
	I_{\text{syn}}(V(t),t) = -g_E(t) (V-E_E) - g_I(t)(V-E_I) = - I_{\text{E}} - I_{\text{I}}. 
\end{align}
In \cite{So2012}, the authors have used the quantity $I_{\text{GPe,STN}}$ in the STN model. However, we know that STN-DBS generate
both excitatory and inhibitory postsynaptic potentials in
STN neurons \cite{Chiken2016}. In our current consideration, instead of using the current $I_{\text{GPe,STN}}$, we consider the current $I_{\text{STN,DBS}} = - I_{\text{E}} - I_{\text{I}}$.
%
%
Let us define the following synaptic dynamics of the STN cell membrane potential ($V$) described by the following model (based on \cite{So2012})
\begin{align}\label{main_eq}
	C_m \frac{d}{dt}V(t) &= - I_L - I_{\text{Na}} - I_\text{K} -  I_{\text{T}} - I_\text{Ca} - I_{\text{ahp} } - I_{\text{STN,DBS}} + I_{\text{app}}  + I_{\text{dbs}} \quad \text{ if } V(t) \leq V_{\text{th}}, \\
	V(t) &= V_{\text{reset}} \quad \text{ otherwise},
\end{align}
where $I_{\text{app}}$ is the external input current, while $C_m$ is the membrane capacitance and $t \in [0, T]$ (some fixed time $T$). Additionally, in \eqref{main_eq}, $V_{\text{th}}$ denotes the membrane potential threshold to fire an action potential.  In this model, we assume that a spike takes place whenever $V(t)$ crosses $V_{\text{th}}$ in the STN membrane potential. In that case, a spike is recorded and $V(t)$ resets to $V_{\text{reset}}$ value. Hence, the reset condition is summarized by $V(t) = V_{\text{reset}}$ if $V(t) \geq V_{\text{th}}$. The quantity $I_{\text{ahp} }$ represents the calcium-activated potassium current for the spike after hyperpolarization in STN.

The concentration of intracellular Ca2+ is governed by the following
calcium balance equation
\begin{align}\label{main_eq2}
	\frac{d}{dt}Ca(t) = \varepsilon(I_{\text{Ca} }- I_T - k_\text{Ca}\text{Ca}(t)),
\end{align}
where $\varepsilon = 3.75\times10^{-5}$ is a scaling constant, $k_\text{Ca} = 22.5$ (ms$^{-1}$) is a given time constant (see, e.g., \cite{Traub1999,Cornelisse2000}).  

Furthermore, we consider an external random (additive noise) input current as follows:
$I_{\text{app}} = \mu_{\text{app}}  + \sigma_{\text{app}} \eta(t)$, where $\eta$ is the zero-Gaussian white noise with $\mu_{\text{app}}>0$ and  $\sigma_{\text{app}}>0$. Using the description of such random input current in our system, the first equation \eqref{main_eq} can be considered as the following Langevin stochastic equation (see, e.g., \cite{Roberts2017}):

\begin{align}\label{main_eq02}
	C_m \frac{d}{dt}V(t) &= - I_L - I_{\text{Na}} - I_\text{K} -  I_{\text{T}} - I_\text{Ca}  - I_{\text{ahp}}- I_{\text{E}} - I_{\text{I}} \nonumber\\&+  I_{\text{dbs}} + \mu_{\text{app}} + \sigma_{\text{app}} \eta(t)
	\quad \text{ if } V(t) \leq V_{\text{th}}. 
\end{align}

Therefore, the system \eqref{main_eq}--\eqref{main_eq2} ($t \in [0, T]$) can be rewritten as
\begin{align}
	C_m \frac{d}{dt}V(t) &= - I_L - I_{\text{Na}} - I_\text{K} -  I_{\text{T}} - I_\text{Ca}  - I_{\text{ahp}} - I_{\text{E}} - I_{\text{I}}  \nonumber\\&+  I_{\text{dbs}} + \mu_{\text{app}} + \sigma_{\text{app}} \eta(t)
	\quad \text{ if } V(t) \leq V_{\text{th}}, \label{eq:1} \\V(t) &= V_{\text{reset}} \quad \text{ otherwise}. \label{eq:2}
\end{align}
Furthermore, we consider the following gating variable dynamics (see, e.g., \cite{So2012})
\begin{align}
	\frac{d}{dt}h(t) &= 0.75\frac{h_{\infty}(V) - h(t)}{\tau_h(V)}, \label{eq:3}\\
	\frac{d}{dt}n(t) &= 0.75\frac{n_{\infty}(V) - n(t)}{\tau_n(V)}, \label{eq:4}\\
	\frac{d}{dt}r(t) &= 0.2\frac{r_{\infty}(V) - r(t)}{\tau_r(V)}, \label{eq:5}\\
	\frac{d}{dt}c(t) &= 0.08\frac{c_{\infty}(V) - c(t)}{\tau_c(V)}, \label{eq:6}\\
	\frac{d}{dt}Ca(t) &= \varepsilon(I_{\text{Ca} }- I_T - k_\text{Ca}\text{Ca}(t)). \label{eq:7}
\end{align}
The initial data we use for the system \eqref{eq:1}--\eqref{eq:7} define its initial conditions:

\begin{align}
	V(0) &= V_0,  \label{eq:8}\\
	h(0) &= h_\infty(V_0),  \label{eq:9}\\
	n(0) &= n_\infty(V_0),  \label{eq:10}\\
	r(0) &= r_\infty(V_0),  \label{eq:11} \\
	c(0) &= c_\infty(V_0),  \label{eq:12}\\
	Ca(0) &= \frac{a_\infty(V_0)}{a_\infty(V_0) + b_\infty(V_0)},  \label{eq:13}
\end{align}
where $h_\infty, n_\infty, r_\infty, c_\infty, a_\infty, b_\infty$ are described as in Table \ref{table:1}. 

In our model \eqref{eq:1}--\eqref{eq:13}, as we mentioned above, we use the simplest input spike train with Poisson process in which the stochastic process of interest provides a suitable approximation to stochastic neuronal firings \cite{Teka2014}. The input spikes will be carried out by the quantity $\sum_{k}\delta(t-t_k)$ in the equation \eqref{conductivity} and each input spike arrives independently of the others, meaning that the timing of one spike does not influence the timing of subsequent spikes. The process will be described as follows:
\begin{itemize}
	\item For designing a spike generator of spike train, we define the probability of firing a spike within a short interval  (see, e.g. \cite{Dayan2005}) as $P(1 \text{ spike during } \Delta t) = r_{j}\Delta t$, where $j=e,i$ with $r_e, r_i$ representing the instantaneous excitatory and inhibitory firing rates, respectively.
	\item Then, a Poisson spike train is generated by first subdividing the time interval into a group of short sub-intervals through small time steps $\Delta t$. In our model, we use $\Delta t = 0.1$ (ms).
	\item  We define a random variable $x_{\text{rand}}$ with uniform distribution
	over the range between 0 and 1 at each time step.
	\item  Finally, we compare the random variable $x_{\text{rand}}$ with the probability of firing a spike, which reads:
	
	\begin{align}
		\begin{cases}
			r_j\Delta t > x_{\text{rand}}, \text{ generates a spike},\\
			r_j \Delta t \leq x_{\text{rand}}, \text{ no spike
				is generated}.
		\end{cases}
	\end{align}
\end{itemize}
		%

By using model \eqref{eq:1}--\eqref{eq:13}, we also investigate the effects of random refractory periods. We consider the random refractory periods $t_{\text{ref}}$ as $t_{\text{ref}} = \mu_{\text{ref}} + \sigma_{\text{ref}}\tilde{ \eta}(t)$, where $\tilde{ \eta}(t) \sim \mathcal{N}(0,1)$ is the standard normal distribution, $\mu_{\text{ref}} > 0$ and $\sigma_{\text{ref}} > 0$. 

In this study, we model the external input to neurons using Poisson spike trains, which are commonly employed in computational neuroscience to simulate random background activity \cite{Brunel2000dynamics,Gerstner2002spiking}. Although real neurons show inter-spike intervals that deviate from the Poisson assumption \cite{Shinomoto2003,Maimon2009beyond}, Poisson processes remain a useful tool for isolating the effects of stochastic noise on STDP \cite{Van2001correlation}. Poisson spike trains are simple and effective in providing random, uncorrelated inputs, which allows us to explore the basic dynamics of STDP and noise interactions in the context of PD without introducing additional complexities like input correlations or network-level dynamics.

In general, the information on stimulating activities in a neuron can be provided by the irregularity of spike trains. The time interval between adjacent spikes is called the ISI. The coefficient of variation (CV) of the ISI in a cell membrane potential with multiple inputs can bring useful information about the output of a decoded neuron. In what follows, we will demonstrate that when we increase the value of $\sigma_{\text{ref}}$, the irregularity of the spike trains increases (see also \cite{Gallinaro2021}).  The term spike irregularity refers to the coefficient of variation of interspike intervals, which quantifies the variability in the timing of spikes rather than the variability of the spike count.
%
The spike irregularity of spike trains can be described via the coefficient of variation of the inter-spike-interval (see, e.g., \cite{Christodoulou2001,Gallinaro2021}) as follows:
\begin{align}\label{CV}CV_{\text{ISI}} = \frac{\sigma_\text{ISI}}{\mu_\text{ISI}},\end{align}
where $\sigma_\text{ISI}$ is the standard deviation and $\mu_\text{ISI}$ is the mean of the ISI of an individual neuron. 

In the next section, let us consider the output firing rate as a function of Gaussian white noise mean or direct current value, namely, the input-output transfer function of the neuron.



In our model, we choose the parameter set as in the following Table \ref{table:1}:
\begin{table}[h!]
	\centering
	\caption{Steady-state functions for channel gating variables and time constants for the different ion
		channels (see, e.g., \cite{So2012}).}
	\label{table:1}
		\begin{tabular}{|l||p{3cm}|p{2cm}|p{2cm}|} 
			\hline
			Current  & Gating variables & Gating variables & Parameters \\ [0.5ex] 
			\hline
			$I_{\text{L}} = g_L(v- E_{\text{L}})$ &  &  & $g_L = 2.25$ (nS)\\   &  &  & $E_\text{L} = -60$ (mV)\\ 
			
			$I_\text{Na} = g_\text{Na}m_{\infty}^3(V)h(V)(V- E_{\text{Na}})$ & $m_{\infty} (V) = 1/(1+ \exp(-\frac{V+30}{15}))$ & $h_\infty(V) = 1/(1+ \exp(-\frac{V+39}{3.1})$ & $g_\text{Na} = 37$ \\
			&   &$\tau_h(V) =1+ 500/(1+ \exp(-\frac{V+57}{-3})$  & $E_\text{Na} = 55$ (mV)\\ 
			
			$I_{\text{K}} = g_\text{K}n^4(V)(V - E_\text{K})$ & $n_{\infty} (V) = 1/(1+ \exp(-\frac{V+32}{8}))$ &  & $g_\text{K} = 45$ (nS)  \\&  $\tau_n(V) =1+ 100/(1+ \exp(-\frac{V+80}{-26})$&  & $E_\text{K} = -80$ (mV)\\ 
			
			$I_\text{T} = g_\text{T}a^3_\infty(V)b_\infty^2(r)r(V)(V - E_\text{T})$ & $a_{\infty} (V) = 1/(1+ \exp(-\frac{V+63}{7.8}))$ & $r_\infty(V) = 1/(1+ \exp(\frac{V+67}{2})$ & $g_\text{T} = 0.5$ (nS)\\
			& $b_{\infty} (V) = 1/(1+ \exp(-\frac{V-0.4}{0.1}))$ & $\tau_r(V) =7.1+ 17.5/(1+ \exp(-\frac{V+68}{-2.2})$ & $E_\text{T} = 0$ (mV)\\& $ - 1/(1+ \exp(4))$  &  & \\ 
			
			$I_\text{Ca} = g_\text{Ca}c^2(V)(V- E_{\text{Ca}})$ & $c_{\infty} (V) = 1/(1+ \exp(-\frac{V+20}{8}))$ &  & $g_\text{Ca} = 2$ (nS) \\ &$\tau_c(V) =1+ 10/(1+ \exp(\frac{V+80}{26})$  &  & $E_\text{Ca} = 140$ (mV)\\ 
			
			$I_\text{ahp} = g_\text{ahp}(V- E_\text{ahp})(\frac{\text{Ca}}{\text{Ca} + 15})$ &  &  & $g_\text{ahp} = 20$ (nS) \\ &  &  & $E_\text{ahp} = -80$ (mV)\\ [1ex] 
			
			$I_{\text{dbs}} = 5+5\sin(2\pi t)$ (pA)&  &  & \\ [1ex] 
			\hline
		\end{tabular}
	
\end{table}

Since these parameters have also been used in \cite{So2012} for STN cell membrane potential experiments, we take them for our model validation. Moreover, in our consideration, we use not only the parameters from Table \ref{table:1}, but also the following parameters: $V_{\text{th}} = -55$ (mV), $V_{\text{reset}} = -70$ (mV), $V_0 = -65$ (mV), $\Delta t = 0.1$, $C_m = 10$ (nF), $\tau_{E} =2$ (ms), $\tau_{I} = 5$ (ms), $\bar{g}_E = 1.5$ (nS), $\bar{g}_I = 0.5$ (nS), $r_e = 10$, $r_i = 10$, $n_E = 20$ spike trains, $n_I = 80$ spike trains. Here, $n_E$ and $n_I$ represent the number of excitatory and inhibitory presynaptic spike trains, respectively. 


Mathematically, the developed model \eqref{eq:1}--\eqref{eq:13} is an evolutionary system that combines stochastic differential equations and ordinary differential equations (SDEs-ODEs), where the stochastic membrane potential equation is coupled to the activation and inactivation ion channels equations, as well as to the calcium-activated potassium current equation. This system can be considered as a modified HH system. 

\subsection{Effects of input correlations}

Correlation or synchrony in neuronal activity refers to the relationship between the firing patterns of different neurons, and it can be measured in various ways. In this context, we focus on the spiking activity of neurons. At its simplest, correlation or synchrony refers to the coincident spiking of neurons-that is, when two neurons spike together, they are said to be firing in synchrony or to be correlated \cite{Salinas2000impact}. Neurons can be synchronous in their instantaneous activity, meaning they spike together with some probability. However, synchrony can also occur with a time delay, where the spiking of one neuron at time \( t \) is correlated with the spiking of another neuron at a later time (time-delayed synchrony). Origins of synchronous neuronal activity consist of:

\begin{itemize}
	\item Common inputs: Neurons receiving input from the same sources tend to have correlated activity. The degree of correlation in their inputs influences the degree of correlation in their outputs.
	\item Pooling from correlated sources: Neurons may not share the same input neurons but could receive inputs from other neurons that are themselves correlated.
	\item Direct connections: Neurons connected to each other (either unidirectionally or bidirectionally) can exhibit time-delayed synchrony. Gap-junctions between neurons can also facilitate synchrony.
	\item Similar properties: Neurons with similar intrinsic parameters and initial conditions may also exhibit synchronous behavior.
\end{itemize}

When neurons fire together, their coordinated activity can have a stronger influence on downstream neurons. Synapses are sensitive to the temporal correlations between presynaptic and postsynaptic spikes, and this sensitivity plays a crucial role in shaping functional neuronal networks?essential for unsupervised learning.
While synchrony can reduce the system's dimensionality, strong correlations may sometimes impair the decoding of neuronal activity.
A simple model to study the emergence of correlations involves injecting common inputs into two neurons and measuring the resulting output correlation as a function of the fraction of shared inputs.

In this study, we will investigate how correlations are transferred by calculating the correlation coefficient of spike trains recorded from two unconnected HH neurons that received correlated inputs \cite{Salinas2000impact,Hong2012single}. The input current to HH neurons \( i = 1, 2 \) is given by:
\[
I_i = \mu_i + \sigma_i (\sqrt{1-c} \, \xi_i + \sqrt{c} \, \xi_c),
\]
where \( \mu_i \) represents the temporal average of the current. The Gaussian white noise \( \xi_i \) is independent for each neuron, while \( \xi_c \) is common to all neurons. The parameter \( c \) (where \( 0 \leq c \leq 1 \)) controls the proportion of common and independent inputs, and \( \sigma_i \) is the variance of the total input.
The sample correlation coefficient between the input currents \( I_i \) and \( I_j \) is defined as the sample covariance of \( I_i \) and \( I_j \) divided by the product of the square roots of their sample variances. Specifically, we use the following equations:
\[
r_{ij} = \frac{\mathrm{cov}(I_i, I_j)}{\sqrt{\mathrm{var}(I_i)} \, \sqrt{\mathrm{var}(I_j)}},
\]
\[
\mathrm{cov}(I_i, I_j) = \sum_{k=1}^{L} (I_i^k - \bar{I}_i)(I_j^k - \bar{I}_j),
\]
\[
\mathrm{var}(I_i) = \sum_{k=1}^{L} (I_i^k - \bar{I}_i)^2,
\]
where \( \bar{I}_i \) is the sample mean of \( I_i \), \( k \) is the time bin index, and \( L \) is the total number of samples. Here, \( I_i^k \) represents the current at neuron \( i \) at time \( k \cdot dt \). Note that these formulas for covariance and variance are not fully accurate as they should be divided by \( L-1 \) for sample estimates. However, we omit this correction because it cancels out in the calculation of the sample correlation coefficient.

%

\subsection{STDP in neuromorphic systems and other applications}

With many current and potential applications, STDP is often thought of as an unsupervised brain-like learning mechanism for spiking neural networks (SNNs) that, among other things, has attracted significant attention from the neuromorphic hardware community \cite{Tao2023new,Lu2024deep}. Its ability to mimic biological learning processes makes STDP highly relevant for various applications, including pattern recognition and sensory processing, real-time pattern recognition, stabilized supervised STDP, and synchronization in neural networks. Many models have been proposed to investigate the role of STDP mechanisms across these applications. For example, the human brain is recognized as the most complex entity in the known universe \cite{Herculano2011not}. At the microcircuit level, neuronal cells are organized into layers with various connectivity motifs. While the information processing mechanisms at this level remain not fully understood, investigating these motifs?particularly about STDP?is essential for gaining insights into biological learning mechanisms and the emergence of intelligence. STDP provides plasticity rules that depend on spikes. As such, they are unsupervised learning rules commonly used in spiking SNNs and neuromorphic chips to emulate brain-like information processing. However, a significant performance gap remains between ideal model simulations and their neuromorphic implementations. STDP is implemented in SNNs and neuromorphic chips, serving as an unsupervised learning rule crucial for mimicking brain-like information processing  \cite{Gupta2024unsupervised}. Its applications include noisy spatiotemporal spike pattern detection, which is particularly effective even with low-resolution synaptic efficacy in neuromorphic implementations. This capability enhances performance in various computational tasks, making STDP relevant for advancing neuromorphic hardware \cite{Gutam2021adaptive,Gautam2023adaptive}. STDP is used to train an efficient Spiking Auto-Encoder that leverages asynchronous sparse spikes for input reconstruction, denoising, and classification, achieving superior performance with significantly fewer spikes compared to state-of-the-art methods. 
On the other hand, STDP has significant applications in bio-plausible meta-learning models, particularly in enhancing the adaptability and efficiency of machine-learning systems. By incorporating STDP and Reward-Modulated STDP, these models reflect biological learning mechanisms and enable quick learning in low-data scenarios. This approach is particularly useful for preventing catastrophic forgetting in meta-learning tasks, allowing the model to retain previously acquired knowledge while learning new tasks. Additionally, STDP facilitates the application of these models in spike-based neuromorphic devices, improving their performance in tasks such as few-shot classification and advancing AI systems' capabilities to mimic human-like learning \cite{Khoee2024meta}. STDP is employed to train a Spiking Auto-Encoder that efficiently performs input reconstruction, denoising, and classification with minimal spike usage, showcasing enhanced performance on image datasets while maintaining competitiveness against other learning approaches \cite{Walters2024efficient}. 
STDP is also applied through memristors as synapses to enable in situ learning and inference in SNNs. This approach addresses the computational challenges associated with implementing STDP in hardware, allowing for efficient weight modulation that enhances speed and reduces power consumption. The integration of STDP facilitates real-time pattern recognition by employing a winner-takes-all  mechanism within the SNN architecture. The proposed design significantly improves performance metrics, including power, energy, and accuracy, enabling the classification of 50 million images per second \cite{Hussain2024efficient}. STDP is employed in the Stabilized Supervised STDP learning rule to enhance the classification layer of SNNs, integrating unsupervised STDP for feature extraction and improving performance on image recognition tasks \cite{Goupy2024paired}. Furthermore, STDP is utilized to investigate the configurations needed to achieve robust synchronization in neural networks, with findings that could inform the design of neuromorphic circuits for improved information processing and transmission through synchronization phenomena \cite{Yamakou2023synchronization}. 

STDP can be regarded as a key learning rule in biological neural networks, and its relevance has been explored in neuromorphic systems for a variety of applications. STDP enables synaptic modifications based on the relative timing of pre- and post-synaptic spikes, which is critical for encoding information and adapting to dynamic environments. Recent advancements in flexible neuromorphic electronics have made it possible to integrate artificial synapses and neurons that replicate the functionality of their biological counterparts, providing a platform for computing, soft robotics, and neuroprosthetics. These systems, which emulate synaptic behaviors and exhibit learning capabilities, are particularly promising for future applications in health monitoring and the Internet of Things \cite{Park2020flexible}. Furthermore, neuromorphic computing algorithms, which draw inspiration from biological learning rules like STDP, are becoming central to the development of more efficient, adaptive, and scalable computing technologies. The ongoing research in neuromorphic hardware and algorithms is setting the stage for new opportunities in both machine learning and real-world applications such as cognitive computing and neuroprosthetics \cite{Schuman2022}.

Overall, despite the challenges in scaling STDP for deeper networks and larger tasks, its biological relevance and versatility highlight its significance in advancing artificial intelligence and neural computation. STDP holds the potential to enhance the performance of robotic systems by allowing them to learn from their environments in real time, reinforcing its critical role in the development of intelligent systems.

\section{Numerical results}

In this section, we take a single STN neuron and study how the neuron behaves under random inputs and when it is bombarded with both excitatory and inhibitory spike trains together with the influence of STDP.  The numerical results reported in this section have been obtained using a discrete-time integration based on the Euler method implemented in Python.

In particular, we use the coupled SDEs-ODEs system (2.8)?(2.20) that describes the dynamics of the STN membrane potential. As we have mentioned in the previous section, we will focus on the effects of Gaussian white noise input current together with the random refractory periods on the STN cell membrane potential.

\begin{figure}[h!]
	\centering
	\includegraphics[width=0.8\textwidth]{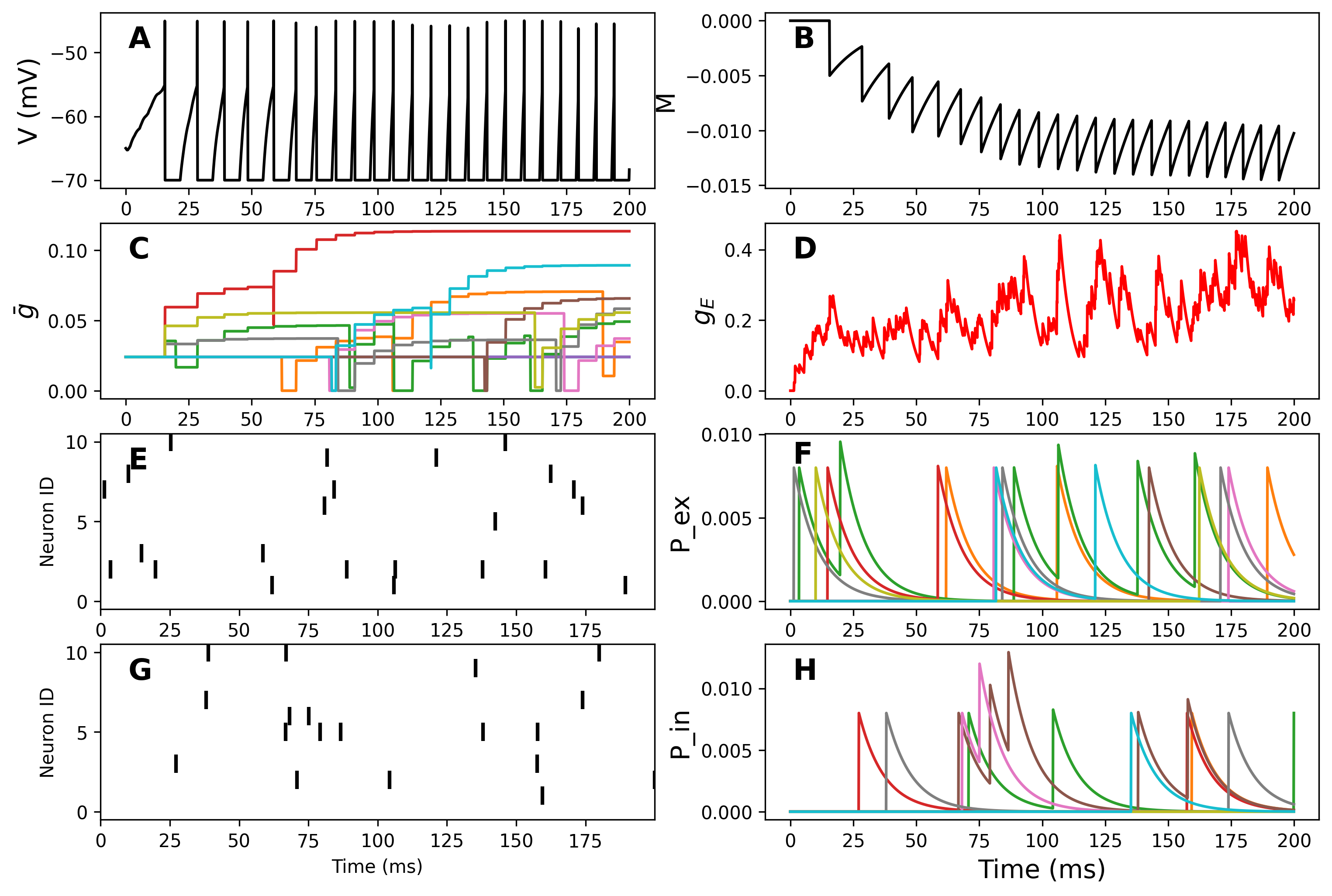} 
	\caption{[Color online] 
		Figures showing the evolution of synaptic conductance and membrane potential over time. (A): Membrane potential (V) as a function of time (in milliseconds) \eqref{main_eq02}. (B): Presynaptic timing (M) dynamics \eqref{post}. (C): Average excitatory synaptic conductance ($\bar{g}$) updates for multiple synapses \eqref{g_M}. (D): Excitatory synaptic conductance ($g_E$) \eqref{conductivity}. (E): Example of presynaptic spike trains for excitatory neurons. (F): The effective conductance or strength of synaptic transmission for excitatory neurons ($P_{ex}$) \eqref{pre}. (G): Example of presynaptic spike trains for inhibitory neurons. (H): The effective conductance or strength of synaptic transmission for inhibitory neurons ($P_{in}$) \eqref{pre}. These panels illustrate the time-dependent changes in membrane potential and synaptic activity profile of direct input current and direct refractory period of healthy cells.}
	\label{fig:0-2}
\end{figure}

\begin{figure}[h!]
	\centering
	\includegraphics[width=0.8\textwidth]{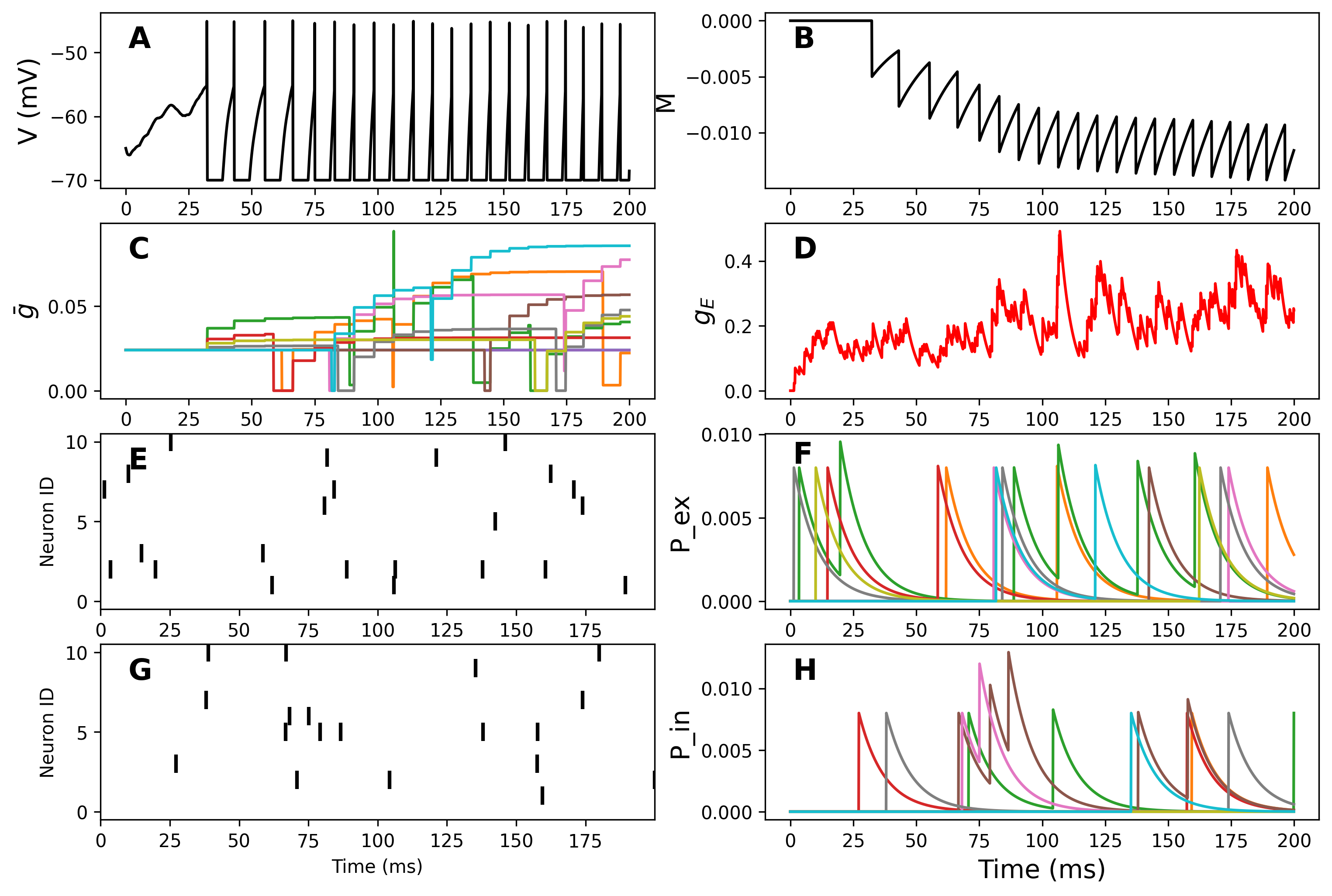} 
	\caption{[Color online] 
		Figures showing the evolution of synaptic conductance and membrane potential over time. (A): Membrane potential (V) as a function of time (in milliseconds)  \eqref{main_eq02}. (B): Postsynaptic timing (M) dynamics \eqref{post}. (C): Average excitatory synaptic conductance ($\bar{g}$) updates for multiple synapses \eqref{g_M}. (D): Excitatory synaptic conductance ($g_E$) \eqref{conductivity}. (E): Example of presynaptic spike trains for excitatory neurons. (F): The effective conductance or strength of synaptic transmission for excitatory neurons ($P_{ex}$) \eqref{pre}. (G): Example of presynaptic spike trains for inhibitory neurons. (H): The effective conductance or strength of synaptic transmission for inhibitory neurons ($P_{in}$) \eqref{pre}. These panels illustrate the time-dependent changes in membrane potential and synaptic activity profile of direct input current and direct refractory period of PD.}
	\label{fig:0-3}
\end{figure}

\begin{figure}[h!]
	\centering
	\includegraphics[width=0.8\textwidth]{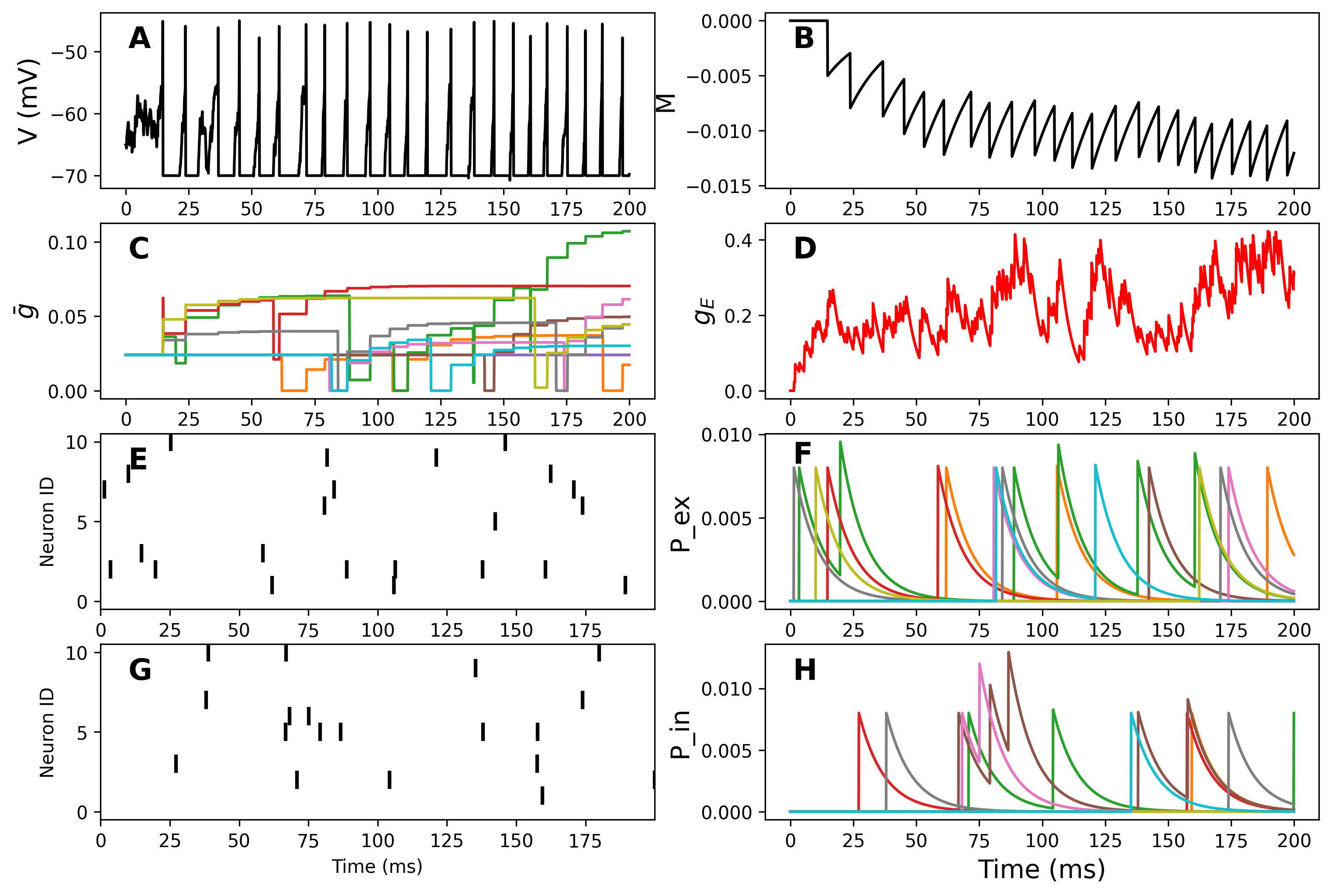} 
	\caption{[Color online] 
		Figures showing the evolution of synaptic conductance and membrane potential over time. (A): Membrane potential (V) as a function of time (in milliseconds)  \eqref{main_eq02}. (B): Postsynaptic timing (M) dynamics \eqref{post}. (C): Average excitatory synaptic conductance ($\bar{g}$) updates for multiple synapses \eqref{g_M}. (D): Excitatory synaptic conductance ($g_E$) \eqref{conductivity}. (E): Example of presynaptic spike trains for excitatory neurons. (F): The effective conductance or strength of synaptic transmission for excitatory neurons ($P_{ex}$) \eqref{pre}. (G): Example of presynaptic spike trains for inhibitory neurons. (H): The effective conductance or strength of synaptic transmission for inhibitory neurons ($P_{in}$) \eqref{pre}. These panels illustrate the time-dependent changes in membrane potential and synaptic activity profile of direct input current and direct refractory period of healthy cells. Parameters: $\sigma = 1, \sigma_{\text{ref}} = 1$.}
	\label{fig:0-4}
\end{figure}

\begin{figure}[h!]
	\centering
	\includegraphics[width=0.8\textwidth]{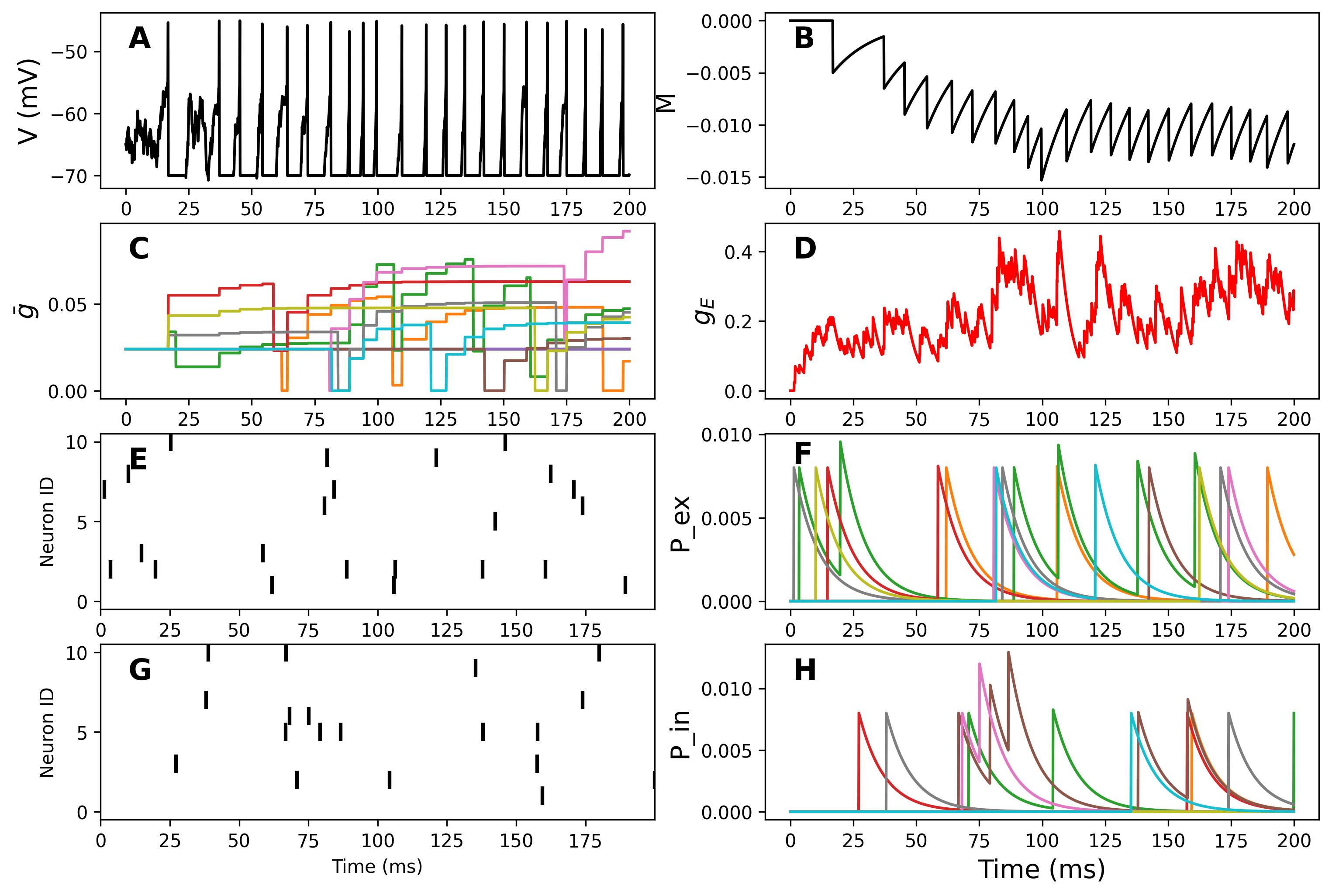} 
	\caption{[Color online] 
		Figures showing the evolution of synaptic conductance and membrane potential over time. (A): Membrane potential (V) as a function of time (in seconds). (B): Presynaptic timing (M) dynamics. (C): Average excitatory synaptic conductance ($\bar{g}$) updates for multiple synapses \eqref{g_M}. (D): Excitatory synaptic conductance ($g_E$) \eqref{conductivity}. (E): Example of presynaptic spike trains for excitatory neurons. (F): Probability of presynaptic spike transmission for excitatory neurons ($P_{ex}$). (G): Example of presynaptic spike trains for inhibitory neurons. (H): Probability of presynaptic spike transmission for inhibitory neurons ($P_{in}$). These panels illustrate the time-dependent changes in membrane potential and synaptic activity profile of random input current and random refractory period of PD. Parameters: $\sigma = 1, \sigma_{\text{ref}} = 1$.}
	\label{fig:0-5}
\end{figure}

\begin{figure}[h!]
	\centering
	\includegraphics[width=0.8\textwidth]{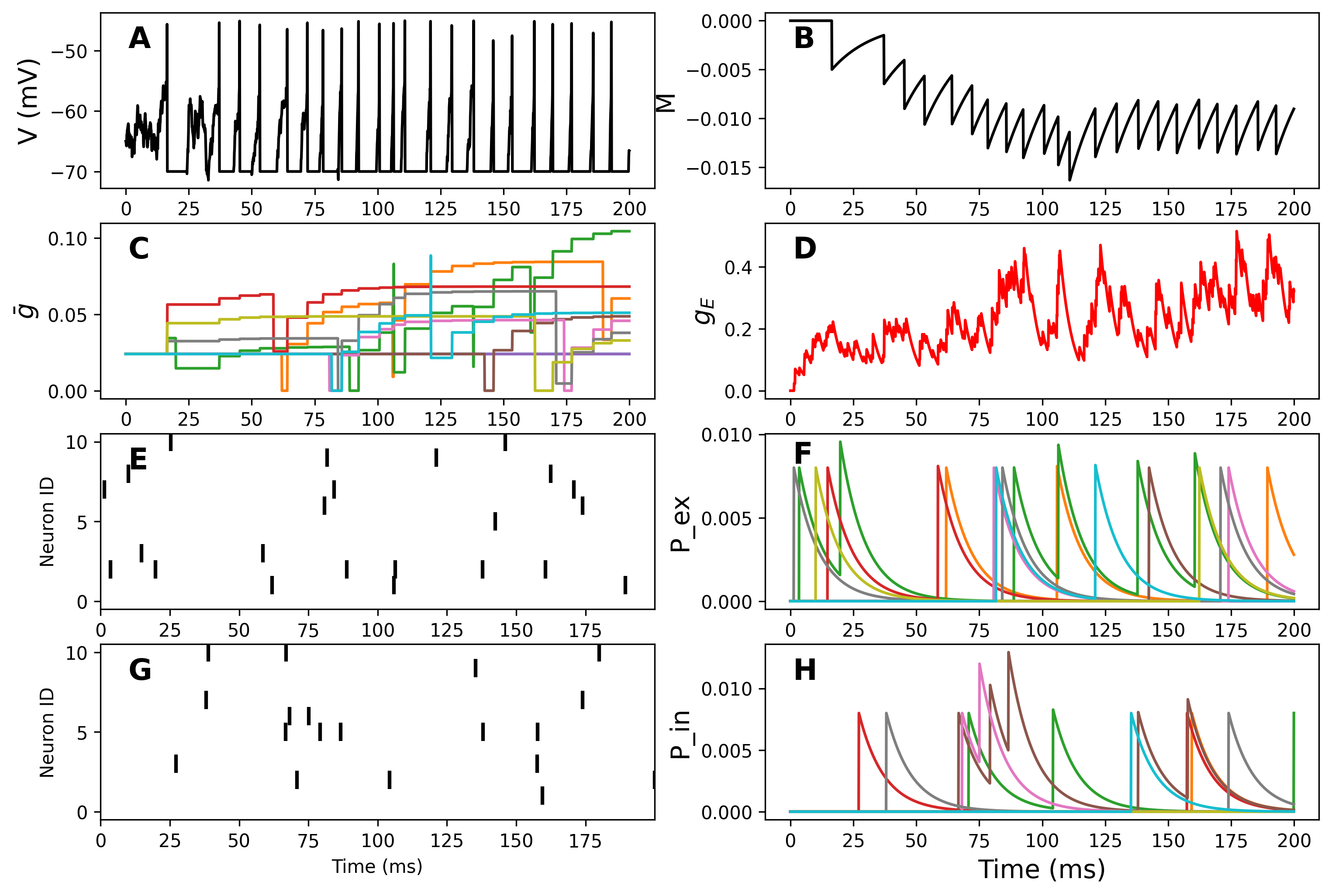} 
	\caption{[Color online] 
		Figures showing the evolution of synaptic conductance and membrane potential over time. (A): Membrane potential (V) as a function of time (in seconds). (B): Presynaptic timing (M) dynamics. (C): Average excitatory synaptic conductance ($\bar{g}$) updates for multiple synapses \eqref{g_M}. (D): Excitatory synaptic conductance ($g_E$) \eqref{conductivity}. (E): Example of presynaptic spike trains for excitatory neurons. (F): Probability of presynaptic spike transmission for excitatory neurons ($P_{ex}$). (G): Example of presynaptic spike trains for inhibitory neurons. (H): Probability of presynaptic spike transmission for inhibitory neurons ($P_{in}$). These panels illustrate the time-dependent changes in membrane potential and synaptic activity profile of random input current and random refractory period of PD cells with DBS.}
	\label{fig:0-6}
\end{figure}

\begin{figure}[h!]
	\centering
	\begin{tabular}{ll}
		\includegraphics[width=0.4\textwidth]{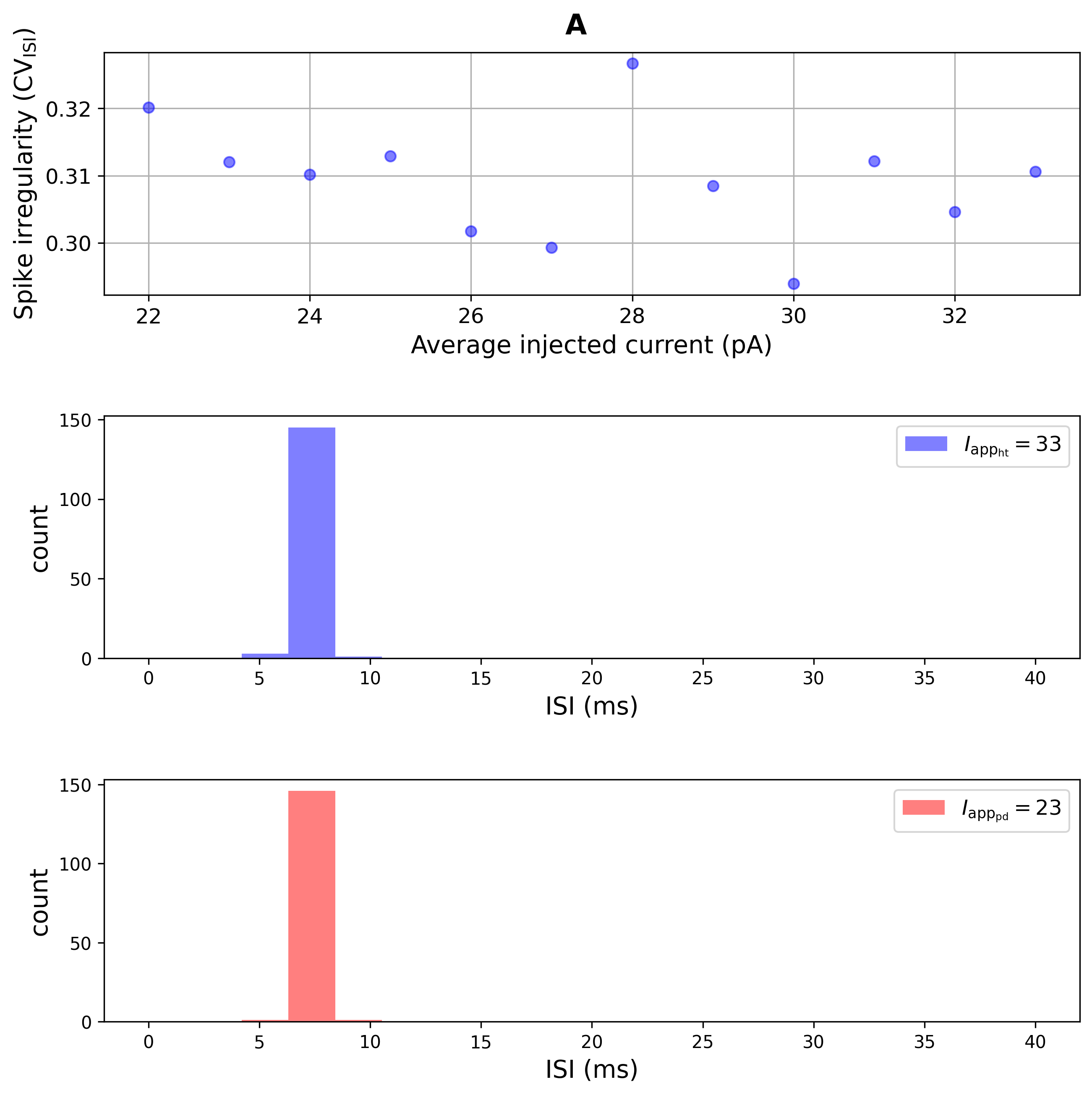}  &
		\includegraphics[width=0.4\textwidth]{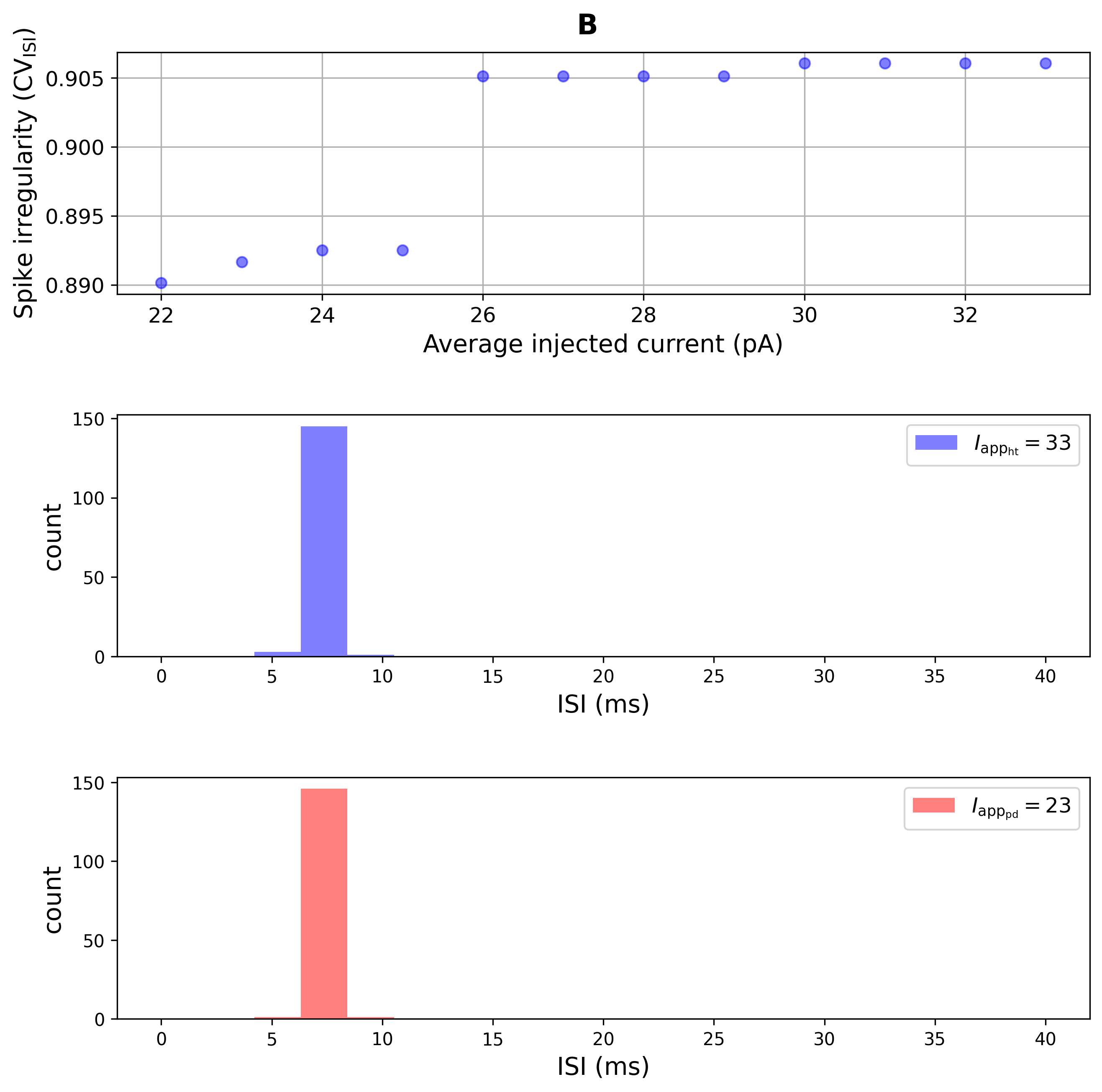}  \\
		\includegraphics[width=0.4\textwidth]{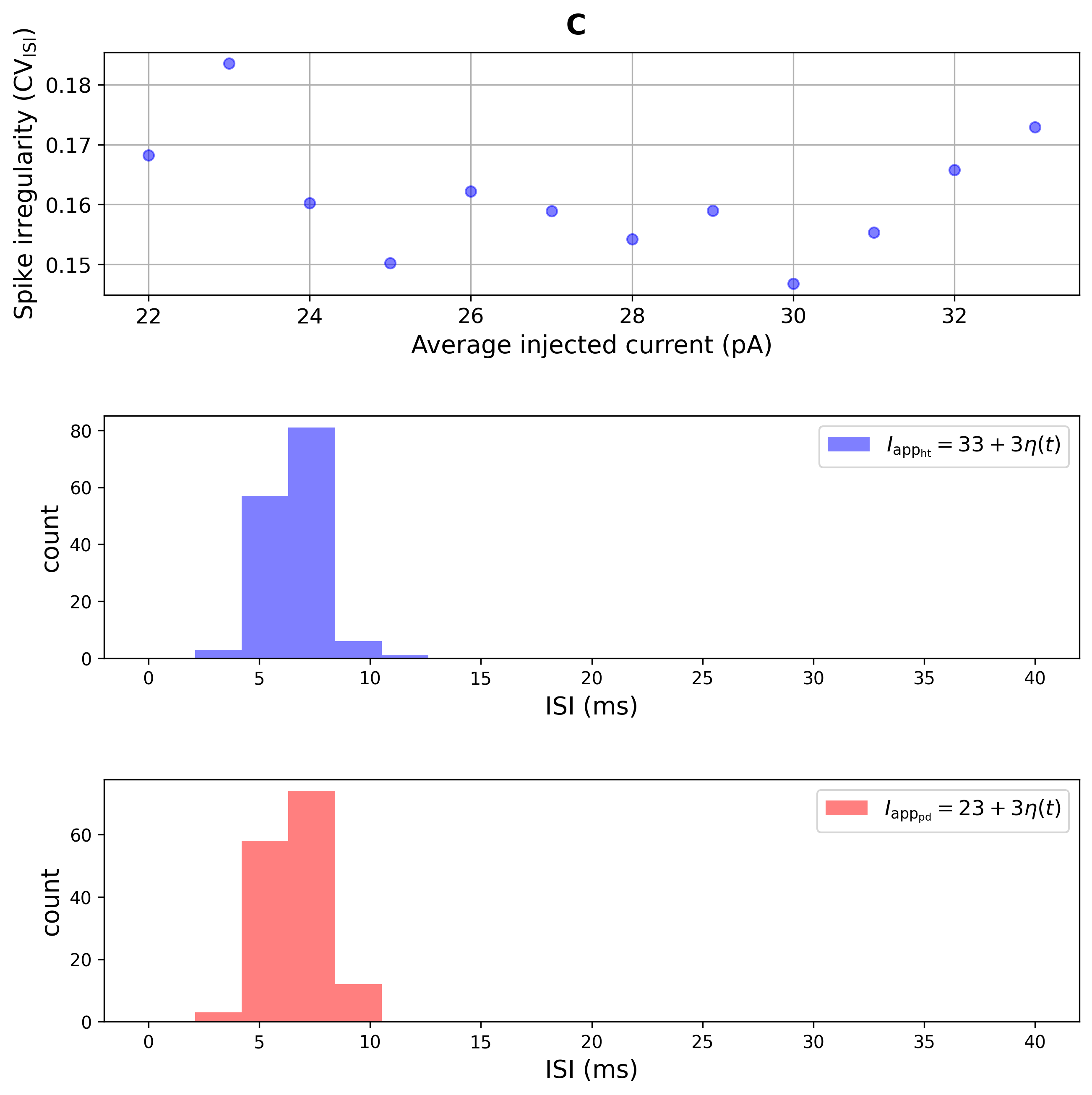}  &
		\includegraphics[width=0.4\textwidth]{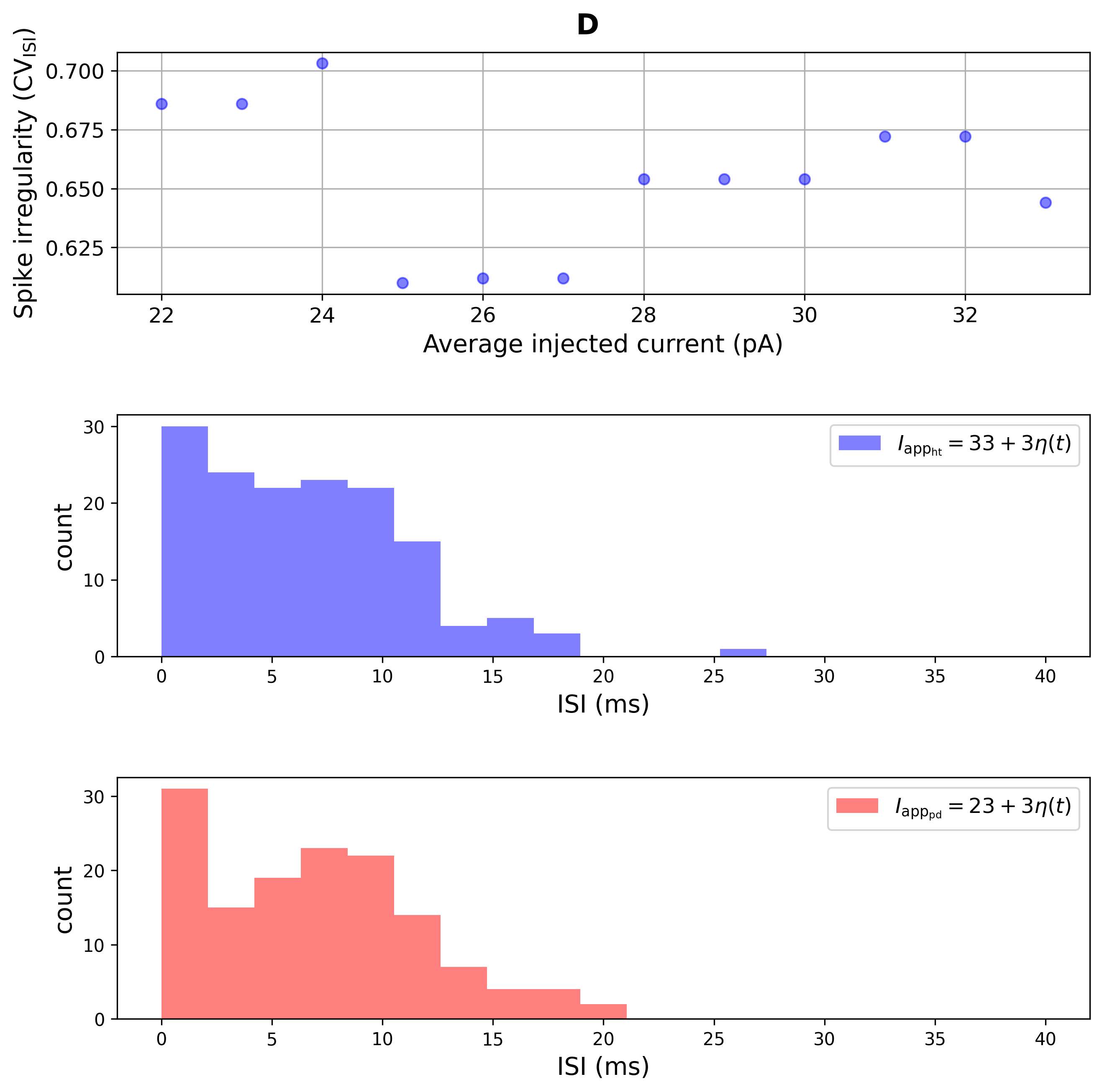}\\
		\includegraphics[width=0.4\textwidth]{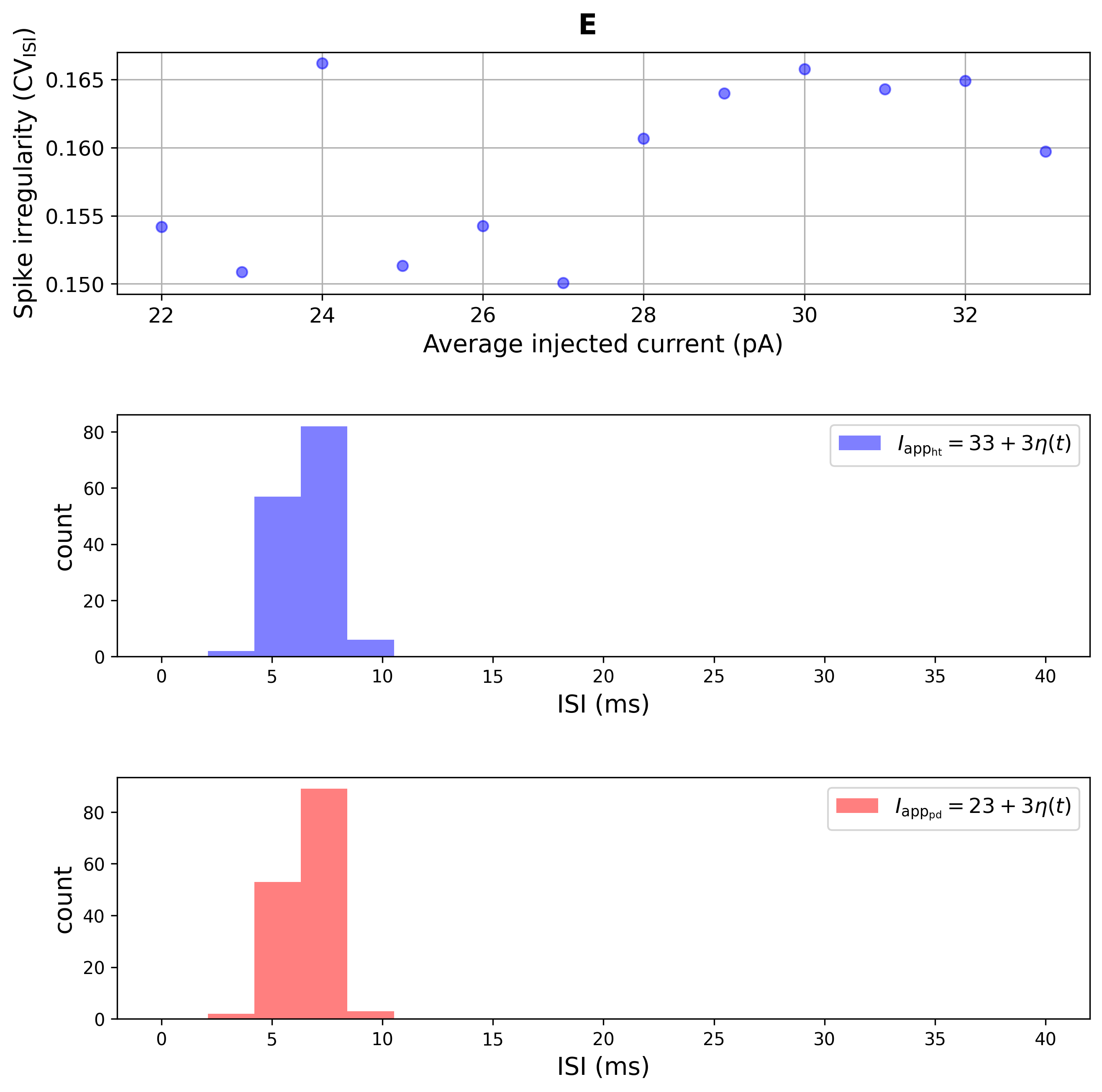}  &
		\includegraphics[width=0.4\textwidth]{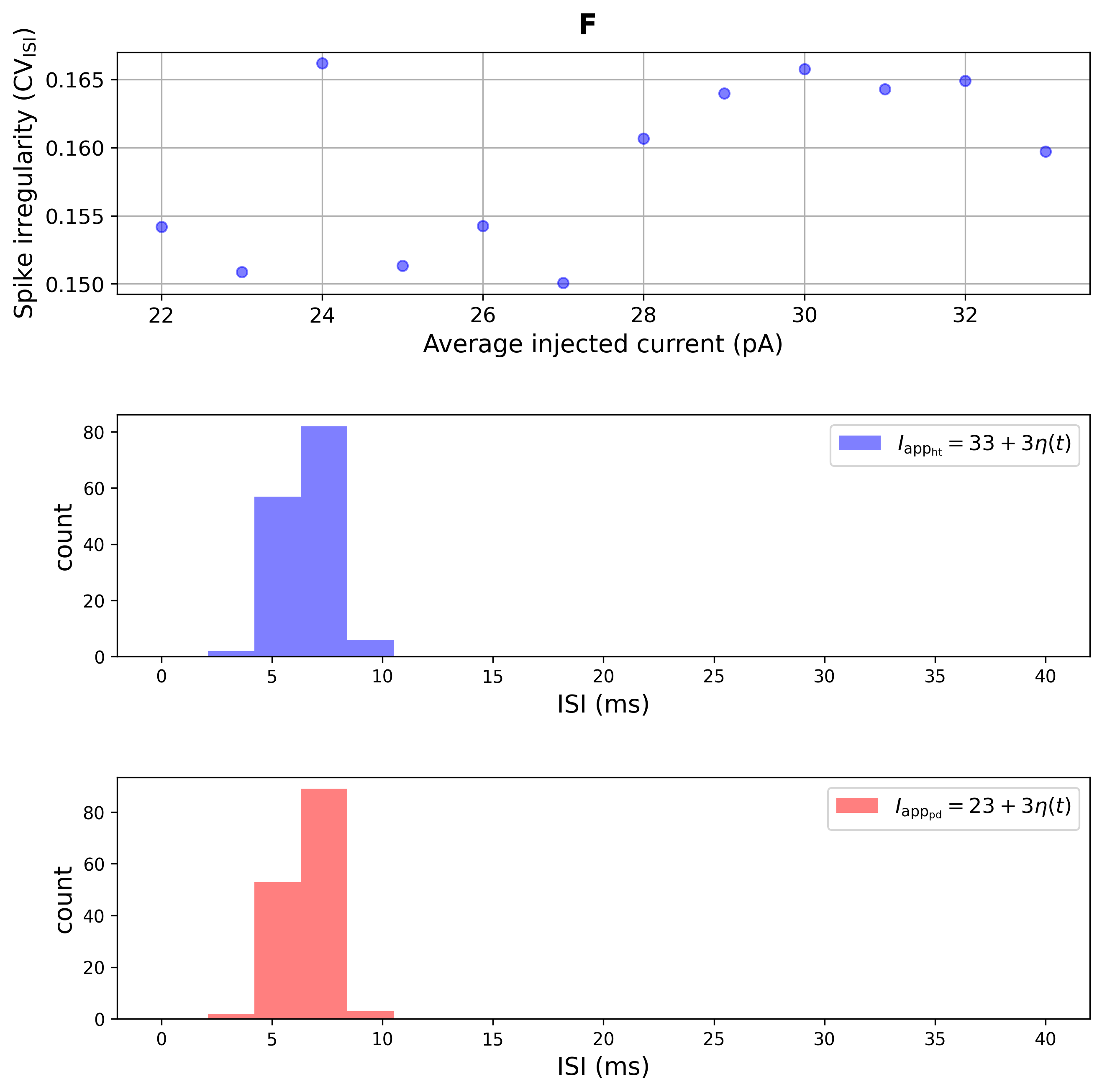}
	\end{tabular}
	\caption{ \small Spike irregularity profiles in the case with the additive noise input current and the ISI distribution. From the top to the bottom, (A) and (B): ISI distributions with direct refractory period. (C) and (D): ISI distributions with random refractory period. (E) and (F): ISI distributions with random refractory period and DBS. Parameters: with $\sigma = 1, \sigma_{\text{ref}} = 1$ in (A), (C) and (E); and $\sigma = 10, \sigma_{\text{ref}} = 5$ in (B), (D) and (F). 
	}\label{fig:sum7}	
\end{figure}

\begin{figure}[h!]
							\centering
							\begin{tabular}{ll}
								\includegraphics[width=0.45\textwidth]{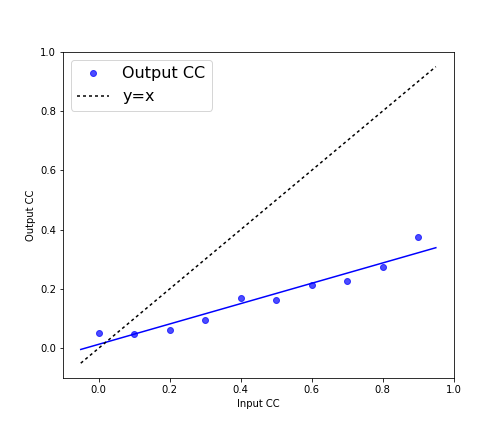} &\includegraphics[width=0.45\textwidth]{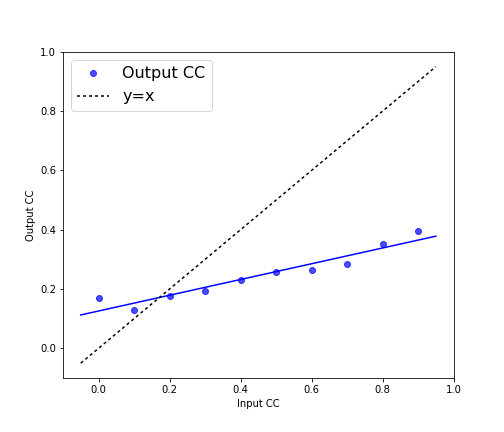} \\
								\includegraphics[width=0.45\textwidth]{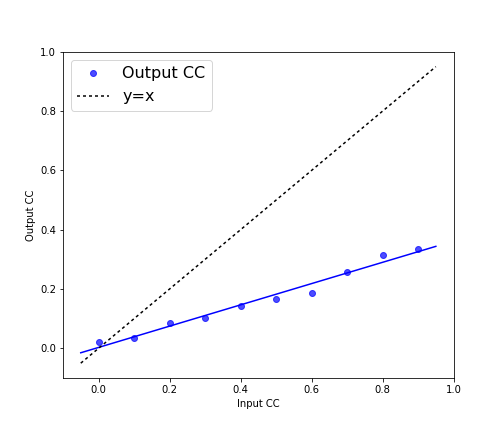} &\includegraphics[width=0.45\textwidth]{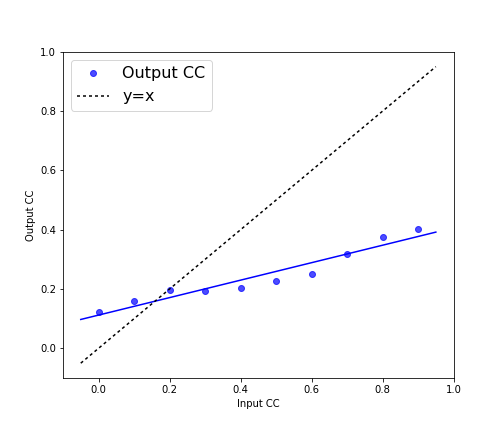}
							\end{tabular}
							\caption{[Color online] Input and output correlations for  PD (first row from the top to the bottom) and healthy cell (second row). First column: $\sigma_{\text{ref}}=1$. Right column: $\sigma_{\text{ref}}=4$.}
							\label{fig:0-17}
						\end{figure}

The main numerical results of our analysis are shown in Figs \ref{fig:0-2} - \ref{fig:0-17}, where we have plotted the time evolution of the membrane potential calculated based on model \eqref{eq:1}-\eqref{eq:13}, along with the spike count profile, the corresponding spike irregularity profile and the effects of input correlations on the output correlations for STN healthy and PD cells with STDP. We investigate the effects of additive type of random input currents in the presence of a random refractory period and the input correlations in a modified HH neuron under synaptic conductance dynamics with STDP. We observe that the spiking activity of a neuron in the STN cell membrane potential is influenced by random external currents, random refractory periods, STDP, and input correlations.

In order to switch from healthy conditions to Parkinsonian conditions in the STN model, we consider a decrease in the current $I_{\text{app}}$ applied to the STN. In particular, we have $I_{\text{app}} = 33$ (pA) for a healthy STN cell and $I_{\text{app}} = 23$ (pA) for a Parkinsonian STN cell (see, e.g., \cite{So2012}). Therefore, an STN cell in the case of injected current input $I_{\text{app}} = 33$ (pA) results in a healthy STN cell, while an STN cell in the case of injected current $I_{\text{app}} = 23$ (pA) is considered as a PD-affected STN cell. 

\begin{figure}[h!]
	\centering
	\includegraphics[width=0.8\textwidth]{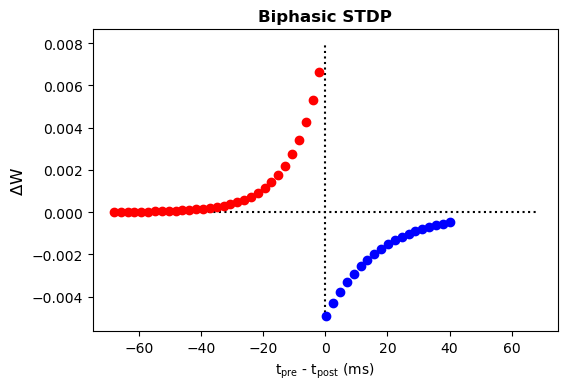} 
	\caption{[Color online] The change in the synaptic weight.}
	\label{fig:0-1}
\end{figure}

In particular, we look at the Fig. \ref{fig:0-2} A, where we have plotted the time evolution of the STN cell membrane potential with direct input current and direct refractory period of healthy cells. The neuron spikes are different between the neuron spikes in the case of healthy cells and with PD cells presented in Figs. \ref{fig:0-2}-\ref{fig:0-3}. There are missing spikes in the time interval of $[0,35]$ in Fig. \ref{fig:0-3} A. 

In Fig. \ref{fig:0-4}, in the case of healthy cell, we added the random input current and random refractory period to the system. We observe that there are fluctuations in the time evolution of the membrane potential (Fig. \ref{fig:0-4} A), the total excitatory synaptic conductance (Fig. \ref{fig:0-4} C) as well as the trace of the number of postsynaptic spikes over the timescale $\tau_{-}$ (Fig. \ref{fig:0-4} B). In the case of PD-affected cells with random input current and random refractory period in Fig. \ref{fig:0-5}, we see that the time evolution of the cell membrane potential is more unstable (Fig. \ref{fig:0-5} A). There are also fluctuations in the trace of the number of postsynaptic conductance (Fig. \ref{fig:0-5} C) and the total excitatory synaptic conductance (Fig. \ref{fig:0-5} B). 

In Fig. \ref{fig:0-6}, we consider a DBS input to the system, there are still fluctuations in the time evolution of the membrane potential (Fig. \ref{fig:0-6} A) and the trace of the number of postsynaptic spikes over the timescale $\tau_{-}$ (Fig. \ref{fig:0-6} B). In what follows, we also look at the corresponding ISI distributions and the spike irregularity profiles of the cases presented in Fig. \ref{fig:sum7}.

In Fig. \ref{fig:sum7}, we present spiking irregularity profiles in the case of an additive noise input current, along with the ISI distribution for healthy and Parkinson?s disease (PD)-affected cells. The variability in ISI is typically measured by its coefficient of variation, $CV_\text{ISI}$. Our analysis focuses on values at $I_\text{app} = 23$ pA and $I_\text{app} = 33$ pA. We first determine the average current injection values, $I_\text{average} = [22, 34]$ pA. The ISI is computed by identifying the spike times and calculating the differences between them. The $CV_\text{ISI}$ is then defined as per equation \eqref{CV}. 

In Fig. \eqref{fig:sum7}, we observe a trend where spike irregularity increases from left to right and decreases from top to bottom. PD cells exhibit greater spike irregularity compared to healthy cells, with a corresponding increase in the ISI range of $CV_\text{ISI}$. It is evident that the presence of a random refractory period significantly influences the spiking activity, reducing irregularity. Additionally, we observe that STDP in the modified HH model increases the ISI of spike trains in output neurons, in contrast to the model without STDP, as reported in \cite{Thieu2022aims}. Moreover, combining a random refractory period with  DBS further reduces spike irregularity.

						Let us combine the two procedures above. First, we generate the correlated inputs. Then we inject the correlated inputs $I_1$ and $I_2$
						into a pair of neurons and record their output spike times. We continue measuring the correlation between the outputs and investigate the relationship between the input correlation and the output correlation.

									The results presented in Fig. \ref{fig:0-17} are obtained by simulating the network response for varying levels of input correlation (see e.g. subsection 2.4). For each input correlation, we compute the output cross-correlation across multiple trials, using a set of predefined model parameters and spike trains for the presynaptic neurons. The plot shows how the input correlation influences the output correlation, reflecting the impact of input correlation on network activity. In Fig. \ref{fig:0-17}, the plot of input correlation versus output correlation is referred to as the correlation transfer function of neurons. The results indicate that output correlation is lower than input correlation and output correlation varies linearly with input correlation.
									
									When input correlations do not affect the neuron?s capacity, output correlation remains independent of both the mean and standard deviation of the refractory period. However, we observe that increasing the standard deviation of the random refractory period in both healthy and PD cells leads to an increase in output correlations, particularly at lower input correlation values (below 0.2). This increase is linear in nature, meaning that as the input correlation increases within this range, the output correlation also increases in a proportional manner. Specifically, this linear trend is most pronounced when input correlations are small, and it reflects how variability in the refractory period amplifies the output correlation as input correlation values approach lower thresholds (see, for example, \cite{De2007correlation,Liu2022effects}).
									
									Moreover, the presence of random inputs influences the spiking activity of STN neurons, both with and without deep brain stimulation (DBS) input currents. As the standard deviation of random refractory periods increases, the irregularity of spike trains decreases. Additionally, random refractory periods can mitigate the effects of random input currents in the system when DBS input currents are present. The interaction between random refractory periods and random input currents in STN cells with STDP membrane potential provides valuable insights for further model developments and progress in DBS therapy.
									
									\section{Conclusions}
									
									We introduced a modified HH model and outlined the process of synaptic conductance with random inputs and STDP. We highlighted the importance of the latter two factors in analyzing and managing neurodegenative diseases, such as Parkinson's, in neuromorphic systems and other applications. By employing a Langevin stochastic dynamics framework in a numerical setting, we investigated the impact of random inputs on the membrane potential of STN cells. Specifically, we detailed the models used and provided numerical examples to examine the effects of random inputs on the time evolution of membrane potentials, neuronal spiking activities, and spike time irregularity profiles. Our results indicate that STDP enhances the regularity of the ISI in spike trains of output neurons. However, the presence of a random refractory period, combined with random input currents, can significantly increase the irregularity of spike trains. Additionally, stochastic influences alongside STDP may enhance the correlation between neurons. These findings could offer insights into managing symptoms of Parkinson's disease.

\backmatter

%
%
%

\bmhead{Acknowledgements}

Authors are grateful to the NSERC and the CRC Program for their
support. RM is also acknowledging support of the BERC 2022-2025 program and Spanish Ministry of Science, Innovation and Universities through the Agencia Estatal de Investigacion (AEI) BCAM Severo Ochoa excellence accreditation SEV-2017-0718.

\section*{Declarations}
\begin{itemize}
	\item Funding
	
	This work was supported by the the NSERC and the CRC Program for their
	support. RM is also acknowledging support of the BERC 2018-2025 program and Spanish Ministry of Science, Innovation and Universities through the Agencia Estatal de Investigacion (AEI) BCAM Severo Ochoa excellence accreditation SEV-2017-0718 and the Basque Government fund AI in BCAM EXP. 2019/00432.
	\item Conflict of interest/Competing interests (check journal-specific guidelines for which heading to use)
	
	There is no conflict of interest.
	\item Ethics approval 
	
	Not applicable 
	\item Consent to participate
	
	Not applicable
	\item Consent for publication
	
	Not applicable
	\item Availability of data and materials
	
	The datasets generated during and/or analysed during the current study are available from the corresponding author on reasonable request.
	\item Code availability 
	
	The codes generated during and/or analysed during the current study are available from the corresponding author on reasonable request.
	\item Authors' contributions
	
	Author 1 designed the research and carried out all simulations, analyzed the data.  Author 1 and Author 2 wrote the article.
\end{itemize}

\bibliography{mybibn-old.bib}

\end{document}